\documentclass[11pt]{article}

     \usepackage{amssymb}
     \usepackage[reqno]{amsmath}
     \usepackage{epsfig} 
\newcommand{\su}{\text{su}}
\newcommand{\ZZ}{\mathbb{Z}}

\def\id{{\rm id}}
\def\cH{{\cal H}} 
\begin{document}
\baselineskip=17pt
\title{\bf D-brane charges on SO(3)}

\author {{\sc Stefan Fredenhagen} \\[2mm]
                           Institut des Hautes Etudes Scientifiques\\
                  F--91440 Bures-sur-Yvette, France}
\vskip.2cm

\date{April 2, 2003}
\begin{titlepage}      \maketitle       \thispagestyle{empty}

\vskip1cm
\begin{abstract}
\noindent 
\normalsize
In this letter we discuss charges of D-branes on the group
manifold SO(3). Our discussion will be based on a conformal
field theory analysis of boundary states in a $\ZZ _{2}$-orbifold
of SU(2). This orbifold differs from the one recently discussed by 
Gaberdiel and Gannon in its action on the fermions and leads to a 
drastically different charge group. We shall consider maximally 
symmetric branes as well as
branes with less symmetry, and find perfect agreement with a recent computation of
the corresponding K-theory groups.
\end{abstract}
\vspace*{-13.9cm}
{\tt {{IHES/P/04/14} \hfill hep-th/0404017}}
\bigskip\vfill
\noindent
{\small e-mail:}{\small\tt stefan@ihes.fr} 
\end{titlepage}

\section{Introduction: The geometric picture}
D-branes on group manifolds and their charges have been an active 
field of research over
the past years. Most of the literature concentrated on the case of
simply-connected group manifolds. Recently, Gaberdiel and Gannon
analysed D-brane charges in orbifolds 
of SU$(n)$~\cite{Gaberdiel:2004yn} which
geometrically describe non-simply connected manifolds having
SU$(n)$ as covering space. In the super-symmetric WZNW model of
SU$(n)$, however, there is not always a unique choice of how to implement the
orbifold. One specific choice has been investigated in the charge analysis
in~\cite{Gaberdiel:2004yn}, the starting point of this letter is a
different choice in the special case of $\text{SO} (3)\sim \text{SU}
(2)/\ZZ _{2}$. We will comment on the generalisation to other groups at the
end of the paper.

\begin{figure}
\begin{center}
\epsfig{file=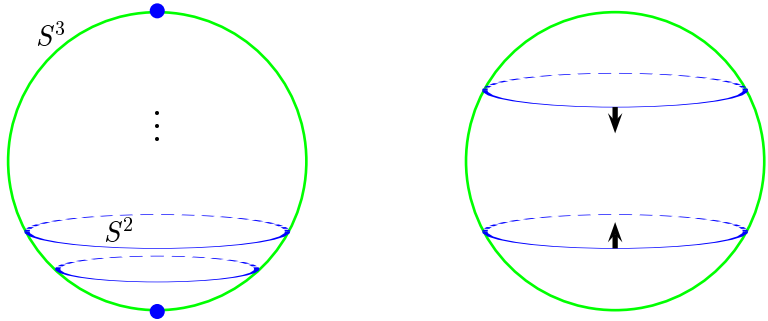}
\end{center}
\caption{\label{figure}The figure on the left sketches the spherical, maximally
symmetric branes in $S^{3}$ in a lower-dimensional illustration. The figure
on the right shows a superposition of oriented branes which are
related by the antipodal map.}
\end{figure}

Geometrically,
SO$(3)$ arises when one identifies antipodal points 
in $S^{3}\sim\  $SU$(2)$. The
maximally symmetric D-branes in SU$(2)$ are localised along conjugacy
classes, 2-spheres $S^{2}$ sitting in $S^{3}$ (see fig.\ \ref{figure}). In the
super-symmetric models, these branes are oriented, and branes of
different orientation correspond to brane and anti-brane, carrying
opposite charges.

Now, D-branes in orbifolds can be described by a superposition of
their preimages in the covering geometry, so maximally symmetric 
D-branes in SO$(3)$ correspond to a superposition of two spherical 
branes in SU$(2)$ related by the antipodal map (for earlier work on
branes in SO$(3)$ 
see \cite{Felder:1999ka,Matsubara:2001hz,Couchoud:2002eg,Bordalo:2003fy}).
From fig.\ \ref{figure} it is obvious that the image of an oriented
$S^{2}$ under the antipodal
map gives back the same $S^{2}$ with the same orientation, translated
from one hemisphere to the other. Therefore we would expect that
the charges of D-branes on SO$(3)$ are given by twice the charges of
branes on SU$(2)$. This does not correspond to the findings of
Gaberdiel and Gannon that the charge inherited from the SU$(2)$-branes
vanishes~\cite{Gaberdiel:2004yn}. Where did we go wrong? What we
tacitly assumed here, is that the
antipodal map acts on the orientation in a natural way. If we instead
combine the antipodal map with a change of orientation, we find that
a brane on SO$(3)$ corresponds to a superposition of brane and
anti-brane on SU$(2)$, and hence should have charge zero in agreement
with~\cite{Gaberdiel:2004yn}.      

The orientation of the branes is related to the fermionic part of the
theory, so the two different implementations of the antipodal map 
should correspond to two conformal field theory descriptions differing
in the way the orbifold acts on the fermions. The first, more natural
action of the antipodal map should correspond to an orbifold which
acts in the same way on bosons and fermions. It is this model we want
to consider here. 

The letter is organised as follows. In the next section we specify our 
model by presenting its field content and analyse the maximally
symmetric branes. The corresponding charge group will be computed in
section 3. In section 4 we shall construct the boundary state of the brane that 
wraps the non-trivial one-cycle of SO$(3)$. We shall discuss the
relation to the recent K-theory computations by Braun and
Sch{\"a}fer-Nameki~\cite{Braun:2004} in the final section 5.

\section{The model and its symmetric branes}
The super-symmetric WZNW model on the simply-connected SU$(2)$ is governed by the
bosonic currents forming an affine algebra $\su(2)_k$ and the fermions which can be combined
into currents of an $\su(2)_2$ algebra. The modular partition function on the torus
is just the diagonal modular invariant of the product theory
$\su(2)_k \otimes \su(2)_2 $ \footnote{Throughout this article, we shall
not distinguish in notation between the chiral algebra and the
underlying affine algebra.}. We can label the sectors by
an integer $l$ running from $0$ to $k$ and an additional label $s$
from the $\su(2)_{2}$ part taking the values $0,1,2$. The values $s=0,2$ 
correspond to the NS sector, $s=1$ to the R sector.  
From the superstring point of view, we can think of the model as a 
GSO-projected theory of type 0 leaving only NSNS and RR sectors. 

The boundary states $|L,S\rangle $ are parametrised by labels with the same 
range because here we are in the 'Cardy case'~\cite{Cardy:1989ir} . 
We can think of the branes with label $S=1$ as an analogue of unstable
'non-BPS' branes (they do not couple to the closed string RR sector), 
and the brane with label $|L,2\rangle $ is the anti-brane to
the one with label $|L,0\rangle $ (their boundary states only differ in the sign
in front of the RR contributions).

The theory we want to consider now is the simple current orbifold of this model
with respect to the simple current $(J_k,J_2)$ of the product theory, i.e.\  the
orbifold acts on fermions as well as on the bosonic
part. Consistency requires the level $k$ to be even. In the notation
of~\cite{Braun:2004}, our model is called the $(+)$-twisted model. 
Note that the $(-)$-twisted model that has been investigated in~\cite{Gaberdiel:2004yn} is
the orbifold with respect to the simple current $(J_{k},\id )$ acting
only on the bosonic part.

The space of states $\cH $ can be obtained by standard techniques (see
e.g.\ \cite{FrancescoCFT}).
For $k=4\ell$, we find (Case A)
\begin{multline*}
\cH \ =\ \bigoplus_{l \text{ even}} \cH _{l}\otimes \overline{\cH }
_{l}\otimes 
(\cH _{0}\otimes \overline{\cH }_{0}\oplus\cH _{2}\otimes
\overline{\cH }_{2} ) \ \oplus \bigoplus _{l \text{
odd}}\cH _{l}\otimes \overline{\cH }_{l}\otimes \cH _{1}\otimes
\overline{\cH }_{1} \\
\oplus  \bigoplus _{l\text{ even}}
\cH_{l}\otimes \overline{\cH }_{k-l}\otimes \cH _{1}\otimes
\overline{\cH }_{1} \ \oplus  \bigoplus_{l\text{
odd}}  \cH_{l}\otimes \overline{\cH }_{k-l}\otimes
(\cH _{0}\otimes \overline{\cH }_{2}\oplus\cH _{2}\otimes
\overline{\cH }_{0}) \ \ ,
\end{multline*}
and for $k=4\ell -2$ (Case B):
\begin{multline*}
\cH \ =\ \bigoplus_{l \text{ even}} \cH _{l}\otimes \overline{\cH }
_{l}\otimes 
(\cH _{0}\otimes \overline{\cH }_{0}\oplus\cH _{2}\otimes
\overline{\cH }_{2} ) \ \oplus \bigoplus _{l \text{
odd}}\cH _{l}\otimes \overline{\cH }_{l}\otimes \cH _{1}\otimes
\overline{\cH }_{1} \\
\oplus  \bigoplus _{l\text{ odd}}
\cH_{l}\otimes \overline{\cH }_{k-l}\otimes \cH _{1}\otimes
\overline{\cH }_{1} \ \oplus  \bigoplus_{l\text{
even}}  \cH_{l}\otimes \overline{\cH }_{k-l}\otimes
(\cH _{0}\otimes \overline{\cH }_{2}\oplus\cH _{2}\otimes
\overline{\cH }_{0}) \ \ .
\end{multline*}

It is well known how to analyse D-branes in orbifold
theories (see e.g.\
\cite{Pradisi:1989xd,Pradisi:1995pp,Douglas:1996sw,
Fuchs:1998xx,Fuchs:1999zi,Birke:1999ik,Behrend:1999bn}),
and we shall explicitly work out the boundary states in our model.
First, we look for Ishibashi states for the untwisted, maximally
symmetric gluing of the SU$(2)$-currents. In both cases, A and B, we find the
states $|l,s\rangle\!\rangle $ for $l+s$ even and $l\not= k/2 $. In
Case A we find in addition the states $|k/2,S\rangle \!\rangle $ for
$S=0,1,2$, in Case B we find the sector $(k/2,1)$ with multiplicity
two, $|k/2,1;+\rangle \!\rangle $ and $|k/2,1;-\rangle \!\rangle $.

Selection rules in the Ishibashi states give rise to identification
rules for the boundary states $|[L,S]\rangle $. The latter are
labelled by equivalence classes of pairs $[L,S]$ with
$[L,S]=[k-L,2-S]$.  
For charge issues we are mainly interested in
boundary states with label $S=0,2$ (see next section). 
We can easily see that in this sector there are no fixed-points under the 
identification, in contrast to 
the $(-)$-twisted model of~\cite{Gaberdiel:2004yn}.

The boundary states for $S=0,2$ are in Case A given by
\begin{equation}\label{bsA}
|[L,S]\rangle \ =\ \sqrt{2}\sum_{l+s\
\text{even}}\frac{S^{(k)}_{Ll}S^{(2)}_{Ss}}{\sqrt{S^{(k)}_{0l}S^{(2)}_{0s}}} |l,s\rangle
\!\rangle \ \ ,
\end{equation}
and in Case B by
\begin{equation}\label{bsB}
|[L,S]\rangle \ =\ \sqrt{2}\!\!\!\! \sum_{\begin{array}{cc}
\scriptstyle l+s\ \text{even}\\[-1mm]
\scriptstyle l\not= k/2
\end{array}}
\!\!\frac{S^{(k)}_{Ll}S^{(2)}_{Ss}}{\sqrt{S^{(k)}_{0l}S^{(2)}_{0s}}} |l,s\rangle
\!\rangle  +
\frac{S^{(k)}_{L\frac{k}{2}}S^{(2)}_{S1}}{\sqrt{S^{(k)}_{0\frac{k}{2}}S^{(2)}_{01}}}
\big(|\tfrac{k}{2},1;+\rangle \!\rangle +|\tfrac{k}{2},1;-\rangle \!\rangle \big)    \ \ .
\end{equation}
Here, $S^{(k)}$ and $S^{(2)}$ are the modular S-matrices of $\su (2)_{k}$
and $\su (2)_{2}$, respectively.

The Cardy computation shows that the NIMreps\footnote{non-negative
integer matrix representations of the fusion ring} for branes with label $S=0,2$
are in both cases
\begin{equation}\label{NIMrep}
n_{(l,s)[L_{1},S_{1}]}{}^{[L_{2},S_{2}]}\ =\
N^{(k)}_{lL_{1}}{}^{L_{2}}N^{(2)}_{sS_{1}}{}^{S_{2}}
+N^{(k)}_{(k-l)L_{1}}{}^{L_{2}}N^{(2)}_{(2-s)S_{1}}{}^{S_{2}} \ \ ,
\end{equation}
where $N^{(k)}$ and $N^{(2)}$ are the fusion matrices of $\su (2)_{k}$
and $\su (2)_{2}$.

In a superstring theory context, the boundary states obtained from a pure
CFT analysis have to be altered 
slightly (see e.g.\ \cite{Recknagel:1998sb,Billo:2000yb}) to obtain correct
GSO projected open string spectra and to respect space-time spin-statistics.
In our case this amounts to an insertion of a factor of $i$ in front of the
RR contributions in the boundary states with $S=0,2$. 
The open string partition function for $S,S'=0,2$ is then
\begin{equation}\label{spectrum}
Z_{[L,S]}{}^{[L',S']} (q)\ =\ \sum_{l,s}n_{(l,s)[L,2-S]}{}^{[L',S']}\, 
\chi^{(k)}_{l}(q)\chi^{(2)}_{s} (q) \ \ ,
\end{equation}
e.g.\ the spectrum of a $[0,0]$-brane is
\[
Z_{[0,0]}{}^{[0,0]} (q)\ =\ \chi^{(k)}_{0}(q)\chi^{(2)}_{2} (q) 
+\chi^{(k)}_{k} (q)\chi^{(2)}_{0} (q)\ \ , 
\]
and does not contain a tachyon.

\section{Charges of symmetric branes}
For the charge analysis we can concentrate on the branes
with label $[L,0]$ and $[L,2]$ (the boundary states with label $S=1$ do not
couple to the RR sector and they always have a tachyon 
in their spectrum). One way of determining brane charges
from the world-sheet point of view is by looking for conserved charges
under RG-flows. In WZNW models, this method has been 
successfully applied in 
many examples
\cite{Alekseev:2000jx,Fredenhagen:2000ei,Bouwknegt:2002bq,Gaberdiel:2003kv}
by considering flows 
described by the Affleck-Ludwig rule~\cite{Affleck:1991by}. In our case, these flows 
give rise to the charge constraint
\[
\dim (l) \, q_{[L_{1},S_{1}]}\ =\
\sum_{[L_{2},S_{2}]}n_{(l,0)[L_{1},S_{1}]}{}^{[L_{2},S_{2}]}
\ q_{[L_{2},S_{2}]} \ \ .
\]
Here, $q_{[L,S]}$ is the abelian charge assigned to $[L,S]$, and $\dim
(l)=l+1$ is the dimension of the corresponding representation of $\su (2)$.
Let us set $S_{1}=0$. From the expression~\eqref{NIMrep}
for the NIMrep we find
\[
\dim (l) \, q_{[L_{1},0]}\ =\
\sum_{L_{2}}N^{(k)}_{lL_{1}}{}^{L_{2}}\ q_{[L_{2},0]}\ \ .
\]
The solution to this charge constraint is
well known from the analysis of SU$(2)$ 
(see e.g.\ \cite{Alekseev:2000jx,Fredenhagen:2000ei}). 
Without further constraints, the charges would take values 
in the finite group $\mathcal{C'}=\ZZ _{k+2}$
and they are given by $q_{[L,0]}=\dim (L)$. This is the same charge
group as for SU$(2)$. We have to be aware of the problem that there 
will be many more flows which are not of the Affleck-Ludwig form, and 
they could restrict the charge group further. 
It actually turns out that not all 
the branes are stable. From~\eqref{spectrum} we see in particular that
the brane with label $L=k/2$ has a tachyon (the vacuum state with $l=0,s=0$) in its 
open string spectrum. Moreover, it does not couple to the closed string RR sector.
Put differently, the brane
$|[\frac{k}{2},0]\rangle $ corresponds in SU$(2)$ to a superposition of 
brane and anti-brane, a chargeless configuration that is not stable. We 
expect that there are RG-flows which lead from a configuration where 
this superposition is present to one where it has vanished; and these
flows should be present also in the SO$(3)$ model. This suggests 
that the brane $|[\frac{k}{2},0]\rangle $ has
charge zero, so $q_{[\frac{k}{2},0]}=\frac{k}{2}+1\stackrel{!}{=}0$. We conclude 
that the charge group of symmetric even-dimensional branes is 
$\mathcal{C}_{\text{even}}=\ZZ _{\frac{k+2}{2}}$. Interestingly, this
result has been anticipated in~\cite{Matsubara:2001hz}.

\section{Symmetry breaking branes and the non-trivial one-cycle}
The branes considered so far preserve as much of the SU$(2)$ symmetry
as possible. Branes with less symmetry in SU$(2)$ have been
constructed in~\cite{Maldacena:2001ky} (see
\cite{Quella:2002ct,Quella:2002ns} for generalisations to other groups). 
The simplest of them corresponds geometrically
to a 1-dimensional circle~\cite{Maldacena:2001ky}. 
In the usual SU$(2)$ model it is not
expected to carry any charge, because there is no non-trivial
one-cycle in the simply connected SU$(2)$. In SO$(3)$ on the other
hand, we would expect that it is possible for a brane to wrap the
non-trivial cycle giving rise to a $\ZZ _{2}$ charge. 

In this section we shall state explicitly the corresponding boundary state. We
identify it by its mass and its localisation region. Under
the assumption that the symmetry breaking branes in SU$(2)$ do not
carry any charges, we show that this brane carries a $\ZZ _{2}$
charge.

What one usually calls 'symmetry breaking boundary states' are in
truth maximally symmetric boundary states with respect to a smaller
symmetry algebra $\mathcal{A}'$. Here, we want to break down the maximal chiral
algebra $\mathcal{A}=\su (2)_{k}\otimes \su (2)_{2}$ in the following way,
\begin{align*}
\mathcal{A}\ =\ \su (2)_{k}\, \otimes\,  \su (2)_{2}\
&\longrightarrow \ \su
(2)_{k}\, \otimes\,  \frac{\su (2)_{2}}{\text{u} (1)_{4}}\, \otimes \,
\text{u} (1)_{4}\\
&\longrightarrow \ \frac{\su (2)_{k}\, \otimes\,  \text{u} (1)_{4}}{\text{u}
(1)_{2k+4}}\, \otimes\,  \text{u}(1)_{2k+4}\, \otimes\,  \frac{\su
(2)_{2}}{\text{u}
(1)_{4}}\  =\ \mathcal{A}'
 \ \ .
\end{align*}
The sectors can be decomposed with respect to $\mathcal{A}'$, 
\[
\cH_{l}\otimes \cH_{s} (q)\ =\ \bigoplus \cH_{[l,s',m]}\otimes  
\cH_{m}\otimes \cH_{[s,s']}  \ \ ,
\]
where $m$ is a $(2k+4)$-periodic integer which labels sectors of $\text{u} (1)_{2k+4}$, and $s'$
is a label of $\text{u} (1)_{4}$ taking the values $s'=0,1,2,3$ modulo $4$.
The coset sectors are labelled by equivalence classes of tuples with 
the identifications $[l,s',m]=[k-l,s'+2,m+k+2]$ and $[s,s']=[2-s,s'+2]$.
The total space of states then reads
\begin{align*}
\cH \  =\ &\bigoplus_{\begin{array}{cc}\scriptstyle l+s\ \text{even}\\[-1mm]
\scriptstyle s_{1}',s_{2}',m_{1},m_{2}
\end{array}} 
\!\!\!\! \cH _{[l,s_{1}',m_{1}]}\otimes \cH _{m_{1}}\otimes \cH _{[s,s_{1}']}
\otimes \overline{\cH}_{[l,s_{2}',m_{2}]}\otimes \overline{\cH
}_{m_{2}}\otimes \overline{\cH }_{[s,s_{2}']}\\
\oplus \!\!\! &\bigoplus_{\begin{array}{cc}\scriptstyle l+s
\begin{array}{ll}\scriptstyle \text{odd (A)}\\[-1.7mm]
\scriptstyle \text{even (B)}
\end{array}\\[-1mm]
\scriptstyle s_{1}',s_{2}',m_{1},m_{2}
\end{array}}
\!\!\! \cH _{[l,s_{1}',m_{1}]}\otimes \cH _{m_{1}}\otimes \cH _{[s,s_{1}']}
 \otimes \overline{\cH}_{[k-l,s_{2}',m_{2}]}\otimes \overline{\cH
}_{m_{2}}\otimes 
\overline{\cH }_{[2-s,s_{2}']} \ \ \ .
\end{align*}
Now we want to implement gluing conditions which twist the currents of
the $u (1)_{2 k+4}$. The closed string sectors $\cH _{\dots }\otimes
\overline{\cH }_{\dots }$ that can couple to such a boundary state are characterised by 
\begin{align*}
[l,s,m]\ &=\ [\bar{l},\bar{s},\bar{m}]\\
m\ &=\ -\bar{m}\\
[s,s']\ &=\ [\bar{s},\bar{s}']\ \ .
\end{align*}
It is straightforward to find the list of Ishibashi states, in Case A
they are labelled by $|l,s,s',m\rangle \!\rangle $ where $m=0,\pm
\frac{k+2}{2},k+2$, and $s+s'$ as well as $l+s+m$ are even; in Case B the states
with $l=\frac{k}{2}$ and $s=1$ occur with multiplicity 2.

For simplicity we only work out the boundary states for Case A. They
are labelled by $|[L,S,S',\hat{M}]\rangle$. Here, $\hat{M}$ can take 
the values $0,1,2,3$ and is defined modulo 4. As always, the selection 
rules on Ishibashi states give rise to identifications among the 
boundary labels, namely 
\[
[L,S,S',\hat{M}]\ = \ [L,2-S,S'+2,\hat{M}]\ = \ 
[k-L, S,S'+2, \hat{M}+2] \ \ .
\]
The boundary state reads explicitly
\[
|[L,S,S',\hat{M}]\rangle \ =\ {\sum}'
\frac{S^{(k)}_{Ll}}{\sqrt{S^{(k)}_{0l}}}
\frac{S^{(2)}_{Ss}}{\sqrt{S^{(2)}_{0s}}}
\frac{Z^{(4)}_{S's'}}{Z^{(4)}_{0s'}}
\frac{\psi_{\hat{M}m}}{Z^{(2k+4)}_{0m}}|l,s,s',m\rangle \!\rangle 
\]
where the sum is restricted to the allowed range for the Ishibashi states. 
We denoted the modular S-matrix of $u(1)_{k}$ by $Z^{(k)}$ to 
distinguish it from the S-matrix of SU$(2)$.
The matrix $\psi$ is defined by 
\[
\psi_{\hat{M}m}\ =\ \frac{1}{2}\exp \big( i\pi m \hat{M}/ (k+2) \big)
\ \ .
\]

Now let us look at charges. As for the symmetric case, all branes can be 
generated from the branes with label $L=0$ \footnote{this can be most
easily seen by employing the rules for boundary RG-flows formulated in
\cite{Fredenhagen:2002qn,Fredenhagen:2003xf}}, so we shall concentrate on these.
The boundary states with label $\hat{M},\hat{M}+2$ arise from a 
fixed-point resolution of symmetry breaking boundary states in SU$(2)$. 
If we assume that these branes in SU$(2)$ do not carry any charge, we 
conclude that the superposition of two branes differing in $\hat{M}$ by 2 
is chargeless. On the other hand, branes with different $\hat{M}$ are 
connected by marginal deformations\footnote{they just correspond to
branes with different Wilson lines along the U$(1)$}, 
and so they have to carry identical 
charges. This means that symmetry-breaking branes in SO$(3)$ can at
most carry a $\ZZ_{2} $ charge. 

The geometric interpretation of these symmetry-breaking branes can be
inferred from the general rules given in
\cite{Maldacena:2001ky,Quella:2002ct}. It stretches along
the image of the $S^{1}$ sitting inside SU$(2)$ and passing through
$-\id $, and hence it wraps the non-trivial one-cycle. We can ask 
whether it really wraps the cycle once. The answer is yes which can be 
seen from the following analysis of masses. 
The length of the cycle is $\pi R$ where
$R\sim \sqrt{k}$ is the radius of the SU$(2)$ (we set $\alpha'=1$). 
The tension of the D1-brane is $\frac{1}{2\pi}$, so the ratio of its mass and 
the mass of the D0-brane (which is described by the boundary state
$|[0,0]\rangle $ of \eqref{bsA}) is  
\[
\frac{\text{Mass of D1-brane on one-cycle}}{\text{Mass of D0-brane}}\
\sim \ \frac{\sqrt{k}}{2}\ \ .  
\]
This should be compared to the ratio of g-factors of the boundary states, 
\[
\frac{g_{|[0,0,0,0]\rangle }}{g_{|[0,0]\rangle }}\
=\ \frac{\psi_{00}}{\sqrt{2}Z^{(2k+4)}_{00}}\ =\ \frac{\sqrt{k+2}}{2}\ \ .
\]
Having in mind that the geometric interpretation applies for large 
values of the level $k$, we find complete agreement.

\section{Discussion}
Let us summarise our results. We analysed brane charges in the
$(+)$-twisted model of SO$(3)$. The charge groups of even-dimensional,
maximally symmetric branes and the contribution of the odd-dimensional
symmetry breaking branes are
\begin{align}\label{cg}
\mathcal{C}_{\text{even}}\ &=\ \ZZ _{\frac{k+2}{2}} &
\mathcal{C}_{\text{odd}}\ &=\ \ZZ _{2} \ \ .
\end{align}

There is much evidence that
D-brane charges are classified by
K-theory~\cite{Minasian:1997mm,Witten:1998cd}. For branes in
backgrounds $X$ with a non-trivial NSNS 3-form field $H$, one has to
use twisted K-groups $^{H}\!K (X)$ \cite{Kapustin:1999di,Bouwknegt:2000qt}. This
applies in particular to WZNW models where the corresponding
K-theories for simply-connected group manifolds have been computed in
\cite{Maldacena:2001xj,Braun:2003rd}. Recently, Braun and
Sch{\"a}fer-Nameki computed the twisted K-theory of SO$(3)$ \cite{Braun:2004}.
The authors found two possible answers
$^{(\pm,H)}K^{*} (\text{SO} (3))$ depending on the choice of a 
twist in $H^{1} (\text{SO} (3),\ZZ _{2})$. One version, $^{(-,H)}K^{*}
(\text{SO} (3))$, is in agreement with the results
of~\cite{Gaberdiel:2004yn} for the $(-)$-twisted model. The other K-groups are 
\begin{align*}
^{(+,H)}K^{1}(\text{SO} (3))\ &=\ \ZZ _{\frac{k+2}{2}} & ^{(+,H)}K^{0}
(\text{SO} (3))\ &=\ \ZZ _{2} \ \ .
\end{align*}
Obviously, these results agree exactly with the charge groups 
$\mathcal{C}_{\text{even}}$ and $\mathcal{C}_{\text{odd}}$ in 
eq.\ (\ref{cg}). 

Let us shortly comment on the role of the symmetry-breaking branes. In
SU$(2)$ and SO$(3)$, these are objects of even co-dimension, and their
charge should be measured by $K^{0}$. In our $(+)$-twisted model, we
found a D1-brane wrapping the non-trivial one-cycle which contributes
a $\ZZ _{2}$-charge in agreement with the K-group $^{(+,H)}K^{0}
(\text{SO} (3))$ above. For the other twist, the result
of~\cite{Braun:2004} is $^{(-,H)}K^{0} (\text{SO} (3))=0$. In the
analysis of the corresponding conformal field theory model
in~\cite{Gaberdiel:2004yn}, symmetry breaking branes have not been
considered, and at first sight one might think that one finds a D1 on
the non-trivial cycle along the same lines as in section 4.
This is not the case. An analysis of masses reveals that a similar 
construction in the $(-)$-twisted model only
gives rise to D1-branes which wrap the cycle twice. The reason is simple: 
in the $(+)$-twisted model, the D1-brane of SU$(2)$ is fixed under 
the antipodal map and gets resolved in SO$(3)$. In the $(-)$-twisted model
on the other hand the D1-brane of SO$(3)$ corresponds to a superposition of 
two D1-branes of different orientations. Therefore, it is 
contractible and we do not expect it to carry any charge; again, we
find agreement with the K-theory result.
\smallskip

This special analysis for $\text{SO} (3)\sim \text{SU} (2)/\ZZ _{2}$
can be generalised to orbifolds of higher rank groups. 
In~\cite{Gaberdiel:2004yn}, brane charges are analysed in the models
$\frac{\su (n)_{k}}{\ZZ _{m}}\otimes \text{so} (d)_{1}$ where $d=n^{2}-1$ is 
the dimension of SU$(n)$. Instead one could analyse the theories
$\frac{\su (n)_{k}\otimes\, \text{so} (d)_{1}}{\ZZ _{m}}$. The action of 
$\ZZ _{m}$ on SU$(n)$ induces a natural action of $\ZZ _{m}$ on SO$(d)$. 
It turns out that this action on the fermions is non-trivial only 
if $n$ is even and $n/m$ is odd, and here we would expect new phenomena.
These are precisely the 'pathological cases' in the analysis
of~\cite{Gaberdiel:2004yn} where the order of the charge group is
unexpectedly small and does not grow with the level $k$. 
Therefore we suggest that one should repeat the 
charge analysis in these cases with the changed orbifold action.  
For comparison, it would be very interesting to compute 
the corresponding K-groups. 

\subsection*{Acknowledgements}
I would like to thank
Pedro Bordalo, Volker Braun, Matthias Gaberdiel, Terry Gannon, Sakura Sch{\"a}fer-Nameki and Volker Schomerus for  
useful discussions and correspondences.

\small
\bibliography{references}
\bibliographystyle{mystyle4}
\end{document}